\newcommand{\nn}{\nonumber\\}

\newcommand{\be}{\begin{eqnarray}}
\newcommand{\ee}{\end{eqnarray}}

\newcommand{\bea}{\begin{eqnarray}}
\newcommand{\eea}{\end{eqnarray}}
\newcommand{\beas}{\begin{eqnarray*}}
\newcommand{\eeas}{\end{eqnarray*}}
\def\gsim{\mathrel{\rlap{\lower4pt\hbox{\hskip1pt$\sim$}}
    \raise1pt\hbox{$>$}}}                
\documentclass[twocolumn,aps,showpacs]{revtex4}
\usepackage{color}
\usepackage{graphics}
\usepackage{epsfig}
\usepackage{amsbsy}
\usepackage{amsmath}
\usepackage{amssymb}
\usepackage{amsthm}
\begin{document}
\title{Dynamical heavy-quark recombination and the non-photonic single
electron puzzle at RHIC}  
\author{Alejandro Ayala$^{1,2}$, J. Magnin$^2$, Luis Manuel Monta\~no$^3$ and
  G. Toledo S\'anchez$^4$}   
\affiliation{$^1$Instituto de Ciencias Nucleares, Universidad
Nacional Aut\'onoma de M\'exico, Apartado Postal 70-543, M\'exico
Distrito Federal 04510, Mexico.\\
$^2$Centro Brasileiro de Pesquisas F\1sicas, CBPF,
Rua Dr. Xavier Sigaud 150, 22290-180, Rio de Janeiro, Brazil.\\
$^3$Centro de Investigaci\'on y de Estudios Avanzados del IPN,
Apartado Postal 14-740, M\'exico Distrito Federal 07000, M\'exico.\\
$^4$Instituto de F\1sica, Universidad Nacional Aut\'onoma de M\'exico,
Apartado Postal 20-364, M\'exico Distrito Federal 01000, Mexico.}

\begin{abstract} 

We show that the single, non-photonic electron nuclear modification factor
$R_{AA}^e$ is affected by the thermal enhancement of the heavy-baryon to
heavy-meson ratio in relativistic heavy-ion collisions with respect to 
proton-proton collisions. We make use of the dynamical quark recombination
model to compute such ratio and show that this produces a sizable suppression
factor for $R_{AA}^e$ at intermediate transverse momenta. We argue that such
suppression  factor needs to be considered, in addition to the energy loss
contribution, in calculations of $R_{AA}^e$.

\end{abstract}


\maketitle

\date{\today}

\section{Introduction}

The suppression of single, non-photonic electrons at RHIC~\cite{PHENIX, STAR}
is usually attributed to heavy-quark energy losses. However,
calculations that successfully describe the nuclear modification factor of
hadrons fail to describe the single, non-photonic electron nuclear
modification factor $R_{AA}^e$~\cite{Djordjevic, Armesto, Wicks}. This has
prompted a great deal of effort aimed to better describe the heavy-quark 
energy loss mechanisms to include not only the radiative 
part~\cite{Baier1, Baier2, Gyulassy3}
but also the collisional~\cite{Djordjevic2} and the medium dynamical
properties to compute the radiative piece~\cite{Djordjevic3}. As a
result, although some improvement in the description of the nuclear
modification factor has been gained, it is not yet clear whether the anomalous
suppression can be completely attributed to energy losses. 

Working along a complementary approach to describe the non-photonic electron
yield at RHIC, it has been argued~\cite{Sorensen, Martinez} 
that under the
assumption of an enhancement in the heavy-quark baryon to meson ratio,
analogous to the case of the proton to pion and the $\Lambda$ to kaon ratios
in Au$ + $Au collisions~\cite{PHENIXBM, STARBM, Laszlo, STAR2},
it is possible to achieve a larger suppression of the nuclear modification
factor. The rationale behind the idea is that heavy-quark mesons have a larger
branching ratio to decay inclusively into electrons as compared to
heavy-quark baryons, and therefore, when the former are less copiously
produced in a heavy-ion environment, the nuclear modification factor decreases,
even in the absence of heavy quark energy losses in the plasma.

In order to give a qualitative argument that shows how an enhancement
in the heavy-quark baryon to meson ratio can suppress the single,
non-photonic electron nuclear modification factor, let us look at the
$p_T$ integrated $R_{AA}^e$ and to consider that the heavy hadrons are only
those containing  a single charm, 
\be
   R_{AA}^{e\ p_T {\mbox{\tiny{int}}}}=\frac{1}{\langle n_p\rangle}
            \frac{N_{AA}^\Lambda B^{\Lambda\rightarrow e}+
                  N_{AA}^D B^{D\rightarrow e}}
            {N_{pp}^\Lambda B^{\Lambda\rightarrow e} + 
            N_{pp}^D B^{D\rightarrow e}},
\label{integratedRAA}
\ee
where $\langle n_p\rangle$ is the average number of participants in
the collision for a given centrality class, $N_{AA\ (pp)}^x$, refers to the
number of $x$-particles produced in A + A (p + p) collisions and
$B^{x\rightarrow e}$ is the branching ratio for the inclusive decay of
$x$-particles into electrons. Let us bring Eq.~(\ref{integratedRAA}) into
a form that contains the corresponding $p_T$ integrated nuclear
modification factor for particles containing charm. We write
\be
   R_{AA}^{e\ p_T {\mbox{\tiny{int}}}}=\frac{1}{\langle n_p\rangle}
            \left(\frac{ N_{AA}^D}{N_{pp}^D}\right)
            \left(\frac{B^{D\rightarrow e} + 
            \frac{N_{AA}^\Lambda}{N_{AA}^D}B^{\Lambda\rightarrow e}}
            {B^{D\rightarrow e} + 
            \frac{N_{pp}^\Lambda}{N_{pp}^D}B^{\Lambda\rightarrow e}}
            \right).            
\label{modintegratedRAA}
\ee
Let us introduce the shorthand notation
\be
   C&=&\frac{N_{AA}^\Lambda / N_{AA}^D}{N_{pp}^\Lambda / N_{pp}^D}\nn
   x&=&\frac{B^{\Lambda\rightarrow e}}{B^{D\rightarrow e}},
\label{shorthand}
\ee
where $C$ represents the {\it enhancement factor} for the ratio of
charm baryons to mesons in A + A as compared to p + p collisions and
$x$ is the charm baryon to meson relative branching ratios for their
corresponding inclusive decays into electrons. With these definitions,
and after rewriting the factor $N_{AA}^D / N_{pp}^D$ in the form
\be
   \frac{N_{AA}^D}{N_{pp}^D}&=&\frac{N_{AA}^D + N_{AA}^\Lambda - N_{AA}^\Lambda}
                                  {N_{pp}^D + N_{pp}^\Lambda - N_{pp}^\Lambda}\nn
   &=&\left(\frac{N_{AA}^D + N_{AA}^\Lambda}{N_{pp}^D +
     N_{pp}^\Lambda}\right)\nn
   &\times&\left(\frac{1 - N_{AA}^\Lambda / (N_{AA}^D + N_{AA}^\Lambda )}
              {1 - N_{pp}^\Lambda / (N_{pp}^D + N_{pp}^\Lambda
		)}\right),
\label{newfactor}
\ee
we can express Eq.~(\ref{modintegratedRAA}) as
\be
   R_{AA}^{e\ p_T {\mbox{\tiny{int}}}}&=&\frac{1}{\langle n_p\rangle}
   \left(\frac{N_{AA}^D + N_{AA}^\Lambda}{N_{pp}^D +
     N_{pp}^\Lambda}\right)\nn
   &\times&\left(\frac{1 - N_{AA}^\Lambda / (N_{AA}^D + N_{AA}^\Lambda )}
              {1 - N_{pp}^\Lambda / (N_{pp}^D + N_{pp}^\Lambda
		)}\right) \nn
   & \times &  \left(\frac{1 + CxN_{pp}^\Lambda / N_{pp}^D}
              {1 + xN_{pp}^\Lambda / N_{pp}^D}\right)\nn
   &\equiv&\frac{1}{\langle n_p\rangle}
   \left(\frac{N_{AA}^D + N_{AA}^\Lambda}{N_{pp}^D +
     N_{pp}^\Lambda}\right)T_{AA}^{e\ p_T {\mbox{\tiny{int}}}}.
\label{newRAA}
\ee
\begin{figure}[t!] 
{\centering
{\epsfig{file=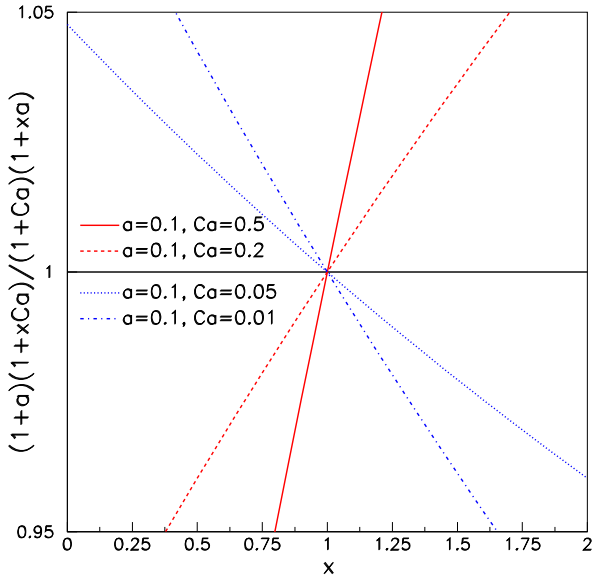, width=1\columnwidth}}
\par}
\caption{(Color online) $p_T$ integrated $T_{AA}^{e}$ as a function of
  $x$, the ratio of 
  branching ratios for charmed baryons and mesons to decay inclusively
  into electrons. Notice that for $x<1$, $T_{AA}^{e}<1$ when $Ca$ --the
  ratio of charm baryons to mesons in A + A-- is larger than $a$ --the
  ratio of charm baryons to mesons in p + p--.}
\label{fig1}
\end{figure}
When not integrated over transverse momentum, the factor $1 / \langle
n_p\rangle \left[(N_{AA}^D + N_{AA}^\Lambda) / (N_{pp}^D +
  N_{pp}^\Lambda)\right]$ represents the nuclear modification factor for
particles with charm.
Let us not assume any particular value for this factor and instead concentrate
in the other one in Eq.~(\ref{newRAA}), which can be written as
\be
   T_{AA}^{e\ p_T {\mbox{\tiny{int}}}}=\frac{(1+a)(1+xCa)}{(1+Ca)(1+xa)},
\label{RAAfinal}
\ee
where $a=N_{pp}^\Lambda / N_{pp}^D$. The above quantity is plotted in
Fig.~\ref{fig1} as a function of $x$ for different combinations of $Ca$ and
$a$. Notice that the function $ T_{AA}^{e\ p_T {\mbox{\tiny{int}}}}$ is
less than 1 when $x < 1$ provided that $Ca > a$.

In this work, we want to quantitatively address the question of whether the
enhancement factor $C$ times $a$ --namely, the heavy-baryon to heavy-meson
ratio in Au + Au collisions-- can indeed be larger than $a$ --namely, the
heavy-baryon to heavy-meson 
ratio in p + p collisions-- and if so, how this affects
the behavior of the factor $T_{AA}^{e}$ as a function of $p_T$. For these
purposes, we use a dynamical recombination scenario that accounts for the fact
that the probability to form baryons and mesons can depend on a different way
on the evolving density during the collision.
 
The work is organized as follows: After presenting a brief introduction to the
dynamical quark recombination model in Sec.~\ref{II}, we proceed in
Sec.~\ref{III} to compute the probabilities to form mesons and
baryons containing a heavy quark in a relativistic heavy-ion collision
environment. In Sec.~\ref{IV} we use these probabilities to write expressions
for the meson and baryon transverse momentum distributions. In
Sec.~\ref{V} we compute such distributions as well as the baryon to meson
ratio. We convolute such ratio with the branching ratios of charmed 
baryons and mesons to decay into electrons to obtain the
$p_T$ unintegrated function $T_{AA}^{e}$ and show that this can be indeed
less than 1. Finally we summarize and conclude in Sec.~\ref{concl}.

\section{Dynamical quark recombination}\label{II}

Recall that hadronization is not an instantaneous process. In fact, lattice
calculations~\cite{Karsch} show that the phase transition from a deconfined
state of quarks and gluons to a hadron gas is, as a function of temperature,
not sharp. Working along this line of thought, it has recently been
shown~\cite{Ayala} that the features of the proton to pion ratio can be well
described by means of the so called {\it dynamical quark recombination model}
(DQRM) that incorporates how the probability to 
recombine quarks into mesons and baryons depends on density and
temperature. Other approaches toward a dynamical description of recombination
have been recently formulated~\cite{Cassing}. 

The upshot of the DQRM is that the density evolving probability differs for
hadrons made up by two and three constituents {\it with the same mass}, that
is to say, the relative population of baryons and mesons can be attributed not
only to flow but rather to the dynamical properties of quark clustering 
in a varying density scenario. A natural question is whether those features
remain true for baryons and mesons with one constituent heavy-quark and
whether a computed, as opposed to assumed, baryon to meson ratio, can at least
partially explain the anomalous single, non-photonic electron suppression at
RHIC. 

The invariant transverse momentum distribution of a given hadron
can be written as an integral over the freeze-out space-time hypersurface
$\Sigma$, of the relativistically invariant phase space particle
density $F(x,P)$,
\be
   E\frac{dN}{d^3P}=g
   \int_{\Sigma_f}d\Sigma\ \frac{P\cdot u(x)}{(2\pi)^3}F(x,P)\, ,
   \label{distribution}
\ee
where $P$ is the hadron's momentum, $u(x)$ is a future oriented unit
four-vector normal to $\Sigma$ and $g$ is the degeneracy factor for the
hadron which takes care of the spin degree of freedom. The function $F(x,P)$
contains the information on the probability that the given hadron is formed. 

To allow for a dynamical recombination scenario in a thermal environment, let
us assume that the phase space particle 
density $F(x,P)$ can be factorized into the product of a term containing the 
thermal occupation number, including the effects of a possible flow
velocity, and another term containing the system energy density $\epsilon$
driven probability ${\mathcal{P}}(\epsilon)$, for the coalescence of partons
into a given hadron. We thus write
\be
   F(x,P)=e^{-P\cdot v(x)/T}{\mathcal{P}}(\epsilon)\, ,
   \label{ourF}
\ee
where $v(x)$ is the flow velocity. As we will show, the probability 
${\mathcal{P}}(\epsilon)$ incorporates in a simple manner the information
that the coalescing partons need to be close in
configuration space as well as to have a not so different velocity. 

To compute the probability ${\mathcal{P}}(\epsilon)$, it has been shown in
Ref.~\cite{Ayala} (where we refer the reader to for details) that use can be
made of the {\it string flip model}~\cite{stringflip, string1, Genaro1} in
order to get information about the likelihood of clustering of constituent
quarks to form hadrons from an effective quark-quark interaction. In short,
the model is a variational quantum Monte Carlo simulation that, taking a set
of equal number of all color quarks and antiquarks at a given density, 
computes the optimal configuration of colorless clusters (baryons or mesons)
by minimizing the potential energy of the system. At low
densities, the model describes the system of quarks as isolated hadrons
while at high densities, this system becomes a free Fermi gas of quarks. The
interaction between quarks is pair-wise and taken as harmonic. The optimal
clustering is achieved by finding the optimal pairing between two given sets
of quarks of different color for all possible color charges.

Consider, for example, two sets with equal amount $A$ of quarks, one  of color
$c_1$ and  
the other of color $c_2$, irrespective of flavor, in accordance to the the
flavor-blindness nature of QCD. We define the optimal pairing between $c_1$
and $c_2$ quarks as the one producing the minimum in the potential energy,
over all possible permutations: 
\be
   V_{c_1 c_2}=\underset{P}{min}\sum\limits_{i=1}^{A}
   \textit{v}[r_{ic_1},P(r_{ic_2})],
   \label{Potencial}  
\ee
where $r_{ic_1}$ is the spatial coordinate of the $i$-th $c_1$ quark
and $P(r_{ic_2})$ is the coordinate of the mapped $j$-th $c_2$ quark.  The
harmonic interaction between pairs is written as
\be
   \textit{v}(r_{ic_1},r_{jc_2})=\frac{1}{2}k(r_{ic_1}-r_{jc_2})^2\, ,
   \label{smalpot}
\ee 
where $k$ is the spring constant. There are two possible kinds of hadrons that
can be formed:  

i) {\it Meson-like}. In this case the pairing is imposed to be between color
and anticolors, and the many-body potential of the system made up of mesons is
given by:
\be
   V_{mes} = V_{B\bar{B}}+V_{G\bar{G}}+V_{R\bar{R}}\,
   \label{mespot}
\ee
where the individual terms are given by Eq.~(\ref{Potencial}) for the
corresponding colors. $R(\bar{R})$, $B(\bar{B})$ and $G(\bar{G})$ are the
labels for red, blue and green color (anticolor) respectively. Note that this
potential can only build pairs.

ii) {\it Baryon-like}. In this case the pairing is imposed to be between the
different colors in all the possible combinations. In this manner, the
many-body potential is:
\be
   V_{bar} = V_{RB}+V_{BG}+V_{RG}\, 
   \label{barpot}
\ee
which can build colorless clusters by linking 3(RBG), 6(RBGRBG),... etc.,
quarks. Since the interaction is pair-wise, the 3-quark clusters are of the
delta (triangular) shape.

According to QCD phenomenology, the formed hadrons should interact
weakly due to the short-range nature of the hadron-hadron interaction. This is
partially accomplished by the possibility of a quark flipping from one cluster
to another. At high energy density, asymptotic freedom demands that quarks
must interact weakly. This behavior is obtained once the average inter-quark
separation is smaller than the typical confining scale. 
 
To describe the evolution of a system of $N$ quarks as a function of the
particle density we consider the quarks moving in a three-dimensional box,
whose sides have length \textit{L}, and the system described by a variational
wave function of the form: 
\be
   \Psi_{\lambda}(\textbf{x}_1,...,\textbf{x}_N)=e^{-\lambda
   V(\textbf{x}_1,...,\textbf{x}_N)}\Phi_{FG}(\textbf{x}_1,...,\textbf{x}_N),
   \label{wavefun}
\ee
where $\lambda$ is the single variational parameter,
$V$(\textbf{x}$_1$,...,\textbf{x}$_N$) is the many-body potential defined in 
Eqs.~(\ref{mespot}) and~(\ref{barpot}) for mesons and baryons respectively,
and $\Phi_{FG}$(\textbf{x}$_1$,...,\textbf{x}$_N$) is the Fermi-gas wave
function given by a product of Slater determinants, one for 
each color-flavor combination of quarks, which are built up of
single-particle wave functions describing a free particle in a box
\cite{Genaro1}.  The square of the variational wave function is the weighting
probability in the sampling, which we carry out using metropolis algorithm. 

The variational wave function is taken to have the form given
in Eq.~(\ref{wavefun}) since we are interested in the evolution of the
system from low to high energy densities. The exponential 
term is responsible of the clustering correlations. At low energy density, the
system is formed by isolated color-singlet hadrons and quarks strongly
interacting inside each cluster; in this case, the exponential term of the
wave function has a big contribution since the average interquark distance is
of the order of the confining scale. In contrast, at high energy density,
where asymptotic freedom takes place, the interaction between quarks is weak
and the system looks like a Fermi gas of quarks. In this case, the inter-quark
separation is much smaller than the confining scale and the effect of the
exponential term vanishes. Notice that these features allow us to identify the
value of the variational parameter $\lambda$ as being directly proportional to
the probability to form a cluster. This fact will be latter exploited to
define the density dependent probability ${\mathcal{P}}(\epsilon)$ since, as
we show below, $\lambda$ changes from a fixed value 
at low density (isolated clusters) to zero at high density (Fermi gas).

\section{Probabilities}\label{III}

All the results we present here come from simulations done with 384 particles,
192 quarks and 192 antiquarks, corresponding to having 32 light quarks and 32
heavy quarks, plus their antiquarks in the three color charges
(anti-charges). Hereafter we refer to light quarks as $u$-quarks
and to the heavy ones as $c$-quarks. To take into account the mass difference
between the $u$ and $c$ quarks we set $m_c=10 m_u$. We have checked that
variations of this particular choice do not affect our relative probabilities.

To determine the variational parameter as a function of density we first
select the value of the particle density $\rho$ in the box, 
which, for a fixed number of particles, means adjusting the box size. Then, we
compute the energy of the system as a function of the variational parameter
using the Monte Carlo method described in the previous section. The minimum of
the energy determines the optimal variational parameter. We repeat the
procedure for a set of values of the particle densities in the region of
interest. To get a measure of the probability to form a cluster, we take the
variational parameter and divide it by its corresponding value at the lowest
density. Notice that since the heavy quarks are not as abundant as the light
ones, they do not contribute to the energy density and thus, within the model,
this last can be computed by assuming that only light flavors contribute.

The information contained in the variational parameter is global in
the sense that it only gives an approximate idea about the average size of the
inter-particle distance at a given density, which is not necessarily the same
for quarks in a single cluster. To correct for this, and in order to find an
appropriate measure of the probability to form baryons and mesons, we need to
multiply these variational parameters by the likelihood to find clusters of
baryons made up of two-light, one-heavy quark and mesons made up of
one-light, one-heavy quark. This likelihood has to consider the fact
that the thermal plasma is mainly made up of light quarks and thus that the
number of produced heavy quarks is relatively small. To accomplish this, notice
that in a model where the interaction between quarks to form clusters is
flavor (as well as color) blind, this likelihood should account only for the
combinatorial probabilities.

Consider the case where one starts out with a set of $n$ $u$-quarks and $m$
$c$-quarks each coming in three colors. The number of possible
colorless baryons containing three quarks of all possible flavors
that can be formed are
\be
   \begin{array}{cc}
   {\mbox{kind}} & {\mbox{number}} \\
       uuu       &      n^3        \\
       uuc       &    3n^2m        \\
       ucc       &    3nm^2        \\
       ccc       &      m^3,
   \end{array}
\label{comb1}
\ee 
and the total number of possible baryons is $(n+m)^3$. The same counting
applies for antibaryons when one starts from the same numbers of antiquarks
instead of quarks. 

Now, consider the case where one starts with a set of $n$ $u$-quarks, $n$
$\bar{u}$-antiquarks, $m$ $c$-quarks and $m$ $\bar{c}$-antiquarks, each coming
in three colors. The number of possible colorless mesons containing
quark-antiquark pairs of all possible flavors that can be formed are
\be
   \begin{array}{cc}
   {\mbox{kind}} & {\mbox{number}} \\
       u\bar{u}  &      3n^2       \\
       u\bar{c}  &      3nm        \\
       \bar{u}c  &      3nm        \\
       \bar{c}c  &      3m^2,
   \end{array}
\label{comb2}
\ee 
and the total number of possible mesons is $3(n+m)^2$. We now ask for the
{\it relative} abundance of baryons with respect to mesons computed under the
above assumptions on the number of light and heavy quarks that we start
from. Since in the case of mesons we are allowing to consider the case
$u\bar{c}$ as well as $\bar{u}c$, we need to include in the counting of
the groups of three quarks also the antibaryons. Thus the relative abundance
is
\be
   \frac{c-{\mbox{baryons}}+c-{\mbox{antibaryons}}}
        {c-{\mbox{mesons}} +c-{\mbox{antimesons}}}
   &=&
   \frac{2\times 3n^2m/(n+m)^3}{2\times nm/(n+m)^2}\nn
   &=&
   \frac{3n}{2(n+m)}.
\label{relative1}
\ee
Let us now impose that the number of $u$-quarks be a multiple $l$ of the
number of $c$-quarks, namely, $n=lm$. Therefore the above relative abundance
can be written as
 \be
   \frac{c-{\mbox{baryons}}+c-{\mbox{antibaryons}}}
        {c-{\mbox{mesons}} +c-{\mbox{antimesons}}}
   =
   \frac{3l}{2(l+1)}.
\label{relative2}
\ee
Notice that in the plasma, the number of $u$-quarks greatly exceeds the number
of $c$-quarks. Therefore a good analytical estimate of the above relative
abundance can be obtained by taking $l\rightarrow\infty$ which gives 
 \be
   \frac{c-{\mbox{baryons}}+c-{\mbox{antibaryons}}}
        {c-{\mbox{mesons}} +c-{\mbox{antimesons}}}
   \stackrel{l\rightarrow\infty}{\longrightarrow}
   \frac{3}{2}.
\label{relative3}
\ee
It can be checked that the asymptotic value 3/2 is rapidly reached, for
instance, by taking $l=30$, the above fraction already becomes $1.475$.

\begin{figure}[t!] 
{\centering
{\epsfig{file=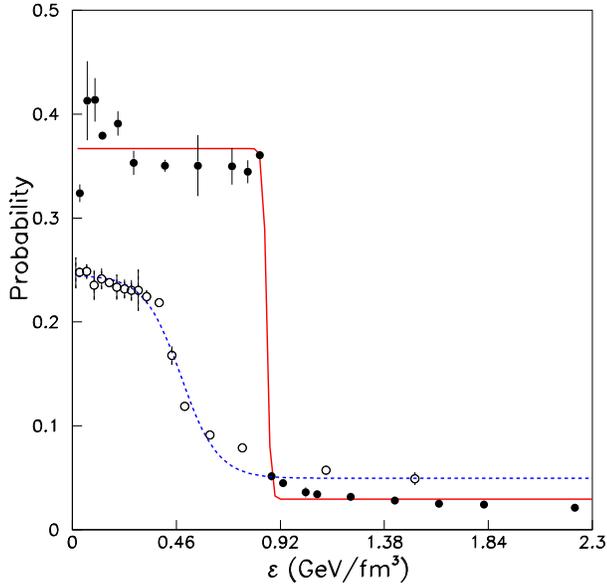, width=1\columnwidth}}
\par}
\caption{(Color online) Probabilities ${\mathcal{P}}^{B,M}$ to produce charmed
  baryons and mesons as a function to the energy density $\epsilon$. Shown are
  the results of the Monte Carlo simulation for baryons (full circles) and
  mesons (open circles) together with a fit to these.}
\label{fig2}
\end{figure}

Figure~\ref{fig2} shows the probability parameter $\cal{P}^{B,M}(\epsilon)$ for 
baryons and mesons, obtained by multiplying the variational parameter with 
the corresponding fraction of baryons/mesons  formed at the given energy 
density. In the case of mesons it corresponds to 1/4 irrespective of the 
density, while for baryons it has a functional form, since the kind of 
clusters can be different as density increases. For low densities the 
ratio of the probabilities becomes 3/2, as expected from the combinatorial 
described above.
Shown in the figure is also a fit to the variational parameters
with the functional form
\be
   f(x)=a_1 + \frac{a_2}{1+\exp{[(x-x_0)/dx]}}. 
\label{fits}
\ee
For baryons
\be
   a_1^B&=& 0.0294\nn
   a_2^B&=& 0.3374\nn
   x_0^B&=& 0.8604\nn
   dx^B&=& 0.0078,
\label{fitbaryon}
\ee
whereas for mesons
\be
   a_1^M&=& 0.0496\nn
   a_2^M&=& 0.1953\nn
   x_0^M&=& 0.4812\nn
   dx^M &=& 0.0813.
\label{fitmeson}
\ee
We will use this analytical expression to carry
out the calculation of the spectra that we proceed to describe.
\begin{figure}[t!] 
{\centering
{\epsfig{file=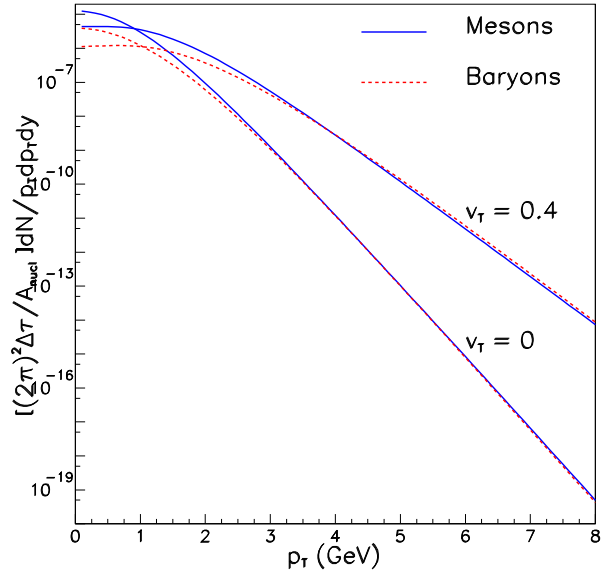, width=1\columnwidth}}
\par}
\caption{(Color online) Charmed baryon and meson transverse momentum
  distributions. The parameters used in the calculation are $m^B=2.29$ GeV,
  $m^M=1.87$ GeV, $\tau_0=1$ fm, $T_0=200$ MeV, $T_f=100$ MeV, corresponding
  to a final time $\tau_f=8$ fm. Shown are the cases with $v_T=0$ and
  $v_T=0.4$.} 
\label{fig3}
\end{figure}

\section{Baryon to meson ratio}\label{IV}

In order to quantify how the different probabilities to produce sets of three
quarks as compared to sets of two quarks affect the
particle's yields as the energy density changes during hadronization, we need
to resort to a model for the space-time evolution of the collision. We take 
Bjorken's scenario which incorporates the fact that initially, expansion is
longitudinal, that is, along the beam direction which we take as the $\hat{z}$
axis and include transverse flow as a small effect on top of the longitudinal
expansion. In this scenario, the relation between the temperature $T$ and the
1+1 proper-time $\tau$ is given by
\be
   T=T_0\left(\frac{\tau_0}{\tau}\right)^{v_s^2},
   \label{temperaturevstau}
\ee
where $\tau=\sqrt{t^2-z^2}$. Equation~(\ref{temperaturevstau}) assumes that
the speed of sound $v_s$ changes slowly with temperature. 
For simplicity we take $v_s$ as a constant equal to the ideal gas limit
$v_s^2=1/3$.

\begin{figure}[!ht] 
{\centering
{\epsfig{file=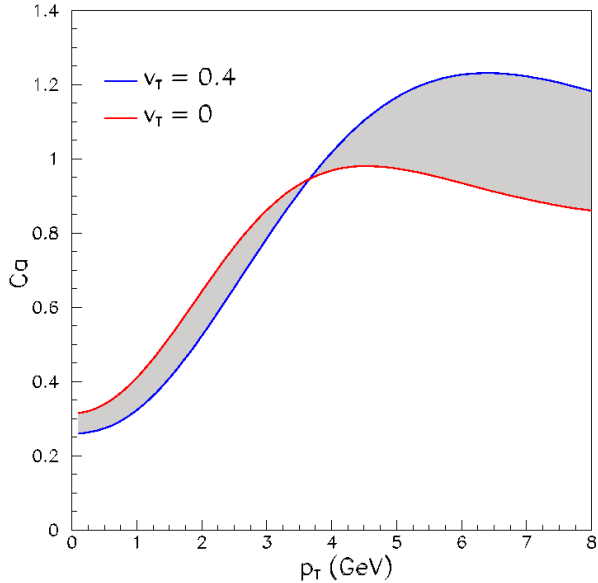, width=1\columnwidth}}
\par}
\caption{(Color online) Charmed baryon to meson ratio, $Ca$, as a function of
  transverse momentum. The parameters used in the calculation are $m^B=2.29$
  GeV, $m^M=1.87$ GeV, $\tau_0=1$ fm, $T_0=200$ MeV, $T_f=100$ MeV,
  corresponding to a final time $\tau_f=8$ fm. Shown is a range when varying
  the transverse expansion velocity $v_T$ from 0 (upper curve at low
  momenta) to 0.4 (lower curve at low momenta).} 
\label{fig4}
\end{figure}
We also consider that hadronization takes place on hypersurfaces $\Sigma$
characterized by a constant value of $\tau$ and therefore
\be
   d\Sigma=\tau\rho \ d\rho \ d\phi\  d\eta ,
   \label{hypersurface}
\ee
where
\be
   \eta=\frac{1}{2}\ln\left(\frac{t+z}{t-z}\right),
   \label{spacerapidity}
\ee
is the spatial rapidity and $\rho$, $\phi$ are the polar transverse
coordinates. Thus, the transverse spectrum for a hadron species $H$
is given as the average over the hadronization interval of the right hand-side
of Eq.~(\ref{distribution}), namely
\be
   E\frac{dN^H}{d^3P}=\frac{g}{\Delta \tau}
   \int_{\tau_0}^{\tau_f}d \tau\int_{\Sigma}d\Sigma\ \frac{P\cdot
     u(x)}{(2\pi)^3}F^H(x,P), 
   \label{distributionaveraged}
\ee
where $\Delta \tau=\tau_f-\tau_0$.

To find the relation between the energy density $\epsilon$ --that the
probability ${\mathcal{P}}$ depends upon-- and $T$, we resort to lattice
simulations. For the case of two flavors (since the heavy quark does not
thermalize), a fair representation of the data~\cite{Karsch} is given by the
analytic expression 
\be
   \epsilon /T^4 = a\left[ 1 + \tanh\left(\frac{T-T_c}{bT_c}\right)\right],
   \label{latticeenergy}
\ee
with $a=4.82$ and $b=0.132$. We take $T_c=175$ MeV.

The flow four-velocity vector $v^\mu$ is given by
\be
   v^\mu&=&(\cosh\eta\cosh\eta_T ,
         \sinh\eta_T\cos\phi ,\nn
        && \sinh\eta_T\sin\phi ,
         \sinh\eta\cosh\eta_T ),
   \label{flowvel}
\ee
where the magnitude of the transverse flow velocity
$v_T$ and $\eta_T$ are related by $v_T = \tanh\eta_T$.
The normal to the freeze-out hypersurfaces of constant $\tau$, $u^\mu$,
is given by
\be
   u^\mu=(\cosh\eta,0,0,\sinh\eta).
   \label{normalhyp}
\ee
We write the momentum four-vector in components as
\be
   P^\mu=(m_T\cosh y,p_T \cos\Phi, p_T \sin\Phi ,m_T\sinh y),
   \label{momvec}
\ee
where $y$ is the $1+1$ momentum rapidity given by
\be
   y=\frac{1}{2}\ln\left(\frac{E+p_z}{E-p_z}\right),
   \label{momentumrapidity}
\ee
and $\Phi$ the azimuthal angle of the momentum components in the transverse
plane. 

Therefore, the products $P\cdot u$ and $P\cdot v$ appearing in
Eq.~(\ref{distributionaveraged}) can be written as
\be
   P\cdot v&=& m_T\cosh (\eta - y)\cos\eta_T -
               p_T\cos (\phi - \Phi )\sinh\eta_T\nn
   P\cdot u&=& m_T\cosh(\eta - y),
   \label{Pdotv}
\ee
Considering the situation of central collisions, we can assume that there is
no dependence of the particle yield on the transverse polar
coordinates. Integration over these variables gives 
\be
   \frac{dN}{p_T dp_T dy}&=&
   g\frac{m_T}{4\pi}\frac{\rho_{\mbox{\tiny{nucl}}}^2}{\Delta\tau}
   \int_{\tau_{0}}^{\tau_{f}}\tau d\tau{\mathcal{P}}(\tau )
   I_0(p_T\sinh\eta_T / T)\nn
   &\times&
   \int d\eta\cosh(y - \eta )e^{-[m_T\cosh(y - \eta )\cosh\eta_T]/T}\nn
   \label{intoverrhophi}
\ee
where $\rho_{\mbox{\tiny{nucl}}}$ is the radius of the colliding nuclei and
$I_0$ is the Bessel function $I$ of order zero. 

We now consider as a further simplification that the space-time and momentum
rapidities are completely correlated, that is $\eta\sim y$. Under this
assumption, the integral over $\eta$ in Eq.~(\ref{intoverrhophi}) can be
performed and we finally get
\be
   \frac{dN}{p_T dp_T dy}&=&
   g\frac{m_T\Delta y}{4\pi}\frac{\rho_{\mbox{\tiny{nucl}}}^2}{\Delta\tau}
   \int_{\tau_{0}}^{\tau_{f}}\tau d\tau{\mathcal{P}}(\tau )\nn
   &\times&I_0(p_T\sinh\eta_T / T)e^{-\cosh\eta_T/T}.
   \label{intoverrhophifin}
\ee
Armed with the expression to compute the hadron transverse momentum
distribution, we  now proceed to apply the analysis to the computation of the
charmed meson and baryon distributions.

\section{Results}\label{V}

Figure~\ref{fig3} shows examples of the transverse momentum distributions
for mesons and baryons obtained from Eq.~(\ref{intoverrhophifin}). We set the
masses of the charmed baryons and mesons as $m^B=2.29$ GeV (corresponding to
$\Lambda_c$) and $m^M=1.87$ GeV (corresponding to $D$). We take the 
initial hadronization time as $\tau_0=1$ fm, at an initial temperature
$T_0=200$ MeV and the final hadronization temperature as $T_f=100$ MeV,
corresponding, according to Eq.~(\ref{temperaturevstau}), to a final time
$\tau_f=8$ fm. Shown are the cases with $v_T=0$ and $v_T=0.4$. Notice
that a finite transverse expansion velocity produces a broadening of the
distributions, as expected.

\begin{figure}[!ht] 
{\centering
{\epsfig{file=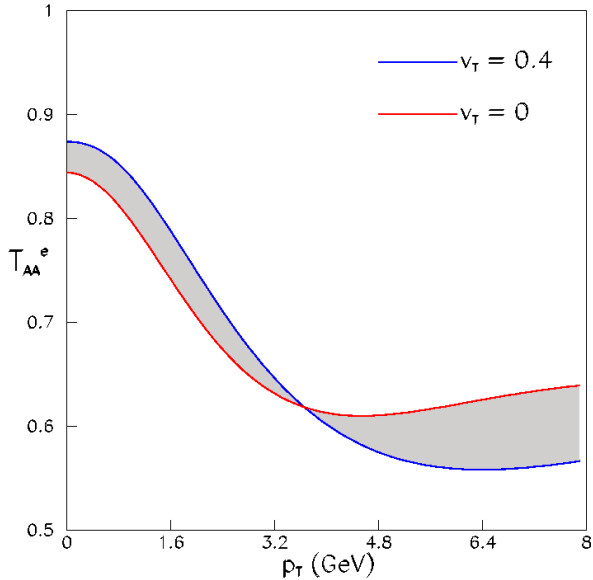, width=1\columnwidth}}
\par}
\caption{(Color online) Suppression factor $T_{AA}^e$ as a function of
  transverse momentum. The parameters used in the calculation are $m^B=2.29$ 
  GeV, $m^M=1.87$ GeV, $\tau_0=1$ fm, $T_0=200$ MeV, $T_f=100$ MeV,
  corresponding to a final time $\tau_f=8$ fm, $x=0.14$, a=0.073. Shown is a
  range for the transverse expansion velocity form $v_T=0$ (upper curve at low
  $p_T$) and $v_T=0.4$ (lower curve at low $p_T$.} 
\label{fig5}
\end{figure}

Figure~\ref{fig4} shows the charmed baryon to meson ratio obtained from the
ratio of the above transverse momentum distributions. Shown is a range for
this ratio when varying the transverse expansion velocity $v_T$ from 0
to 0.4. Notice that for a finite $v_T$, this ratio goes above 1 for
$p_T \gsim 3.5$ GeV. 

We now proceed to compute the $p_T$ unintegrated function
$T_{AA}^{e}$. For this purpose, we take that the possible charmed mesons
decaying inclusively into electrons or positrons are $D^{\pm}$
($B^{D^{\pm}\rightarrow e^{\pm}}=16.0\%$), $D^{0},\bar{D}^{0}$
($B^{D^0,\ \bar{D}^0\rightarrow e^{\pm}}=6.53\%$), $D_s^{\pm}$
($B^{D_s^{\pm}\rightarrow e^{\pm}}=8\%$) and that the possible charmed baryons
decaying inclusively into electrons or positrons are
$\Lambda_c,\bar{\Lambda}_c$ 
($B^{\Lambda_c,\ \bar{\Lambda}_c\rightarrow e^{\pm}}=4.5\%$). Thus we get
\be
   x=0.14.
\label{ratiobranching}
\ee
We also approximate the masses of all the charmed mesons considered to be
equal to the mass of the $D^\pm$ mesons.

From Eq.~(\ref{RAAfinal}) we see that, without integrating over $p_T$, the
dependence on the transverse momentum comes from
$a=(dN_{pp}^\Lambda/dp_T)/(dN_{pp}^D/dp_T)$ and the product 
$Ca=(dN_{AA}^\Lambda/dp_T)/(dN_{AA}^D/dp_T)$. The integrated ratio
$a^{\mbox{\tiny{int}}}$ has been computed in Ref.~\cite{Martinez} using a
Pythia simulation, with the result $a^{\mbox{\tiny{int}}}=0.073$. We have also
performed a simulation using Pythia at NLO with 100,000 events and have found
that with such statistics, the ratio of charmed baryons to charmed mesons in p
+ p collisions at $\sqrt{s_{NN}}=200$ GeV is flat up to $p_T\simeq 5$ GeV and
consistent with the value reported in Ref.~\cite{Martinez}. Therefore, for
simplicity we take $a$ as a constant equal to the above quoted number. 
Thus
\be
   T_{AA}^{e}\simeq \frac{(1+a^{\mbox{\tiny{int}}})}{(1+xa^{\mbox{\tiny{int}}})}
                   \frac{1+x(dN_{AA}^\Lambda/dp_T) / (dN_{AA}^D/dp_T)}
                        {1+(dN_{AA}^\Lambda/dp_T) / (dN_{AA}^D/dp_T)}. 
\label{approxT}
\ee
Figure~\ref{fig5} shows $T_{AA}^{e}$ as a function of $p_T$. We have used
a range of values for the transverse expansion velocity between $v_T = 0$ and
$v_T = 0.4$. We see that for the chosen evolution parameters, $T_{AA}^{e}$ is
indeed smaller than 1 and thus it contributes to the suppression of the single
non-photonic electron nuclear modification factor $R_{AA}^e$.

\section{Conclusions}\label{concl}

In this work we have shown that the anomalous suppression of the single
non-photonic electron nuclear modification factor $R_{AA}^e$ can be partially
understood by realizing that this quantity is affected by an enhancement in
the charmed baryon to meson ratio at intermediate $p_T$ in Au + Au
collisions. This enhancement is due to the fact that in this region, thermal
recombination becomes the dominant mechanism for hadron production. We have
made use of the DQRM to calculate this ratio and have shown that for moderate
and even for vanishing transverse expansion velocities, it indeed can be
larger than the charmed baryon to meson ratio in p + p collisions. This
enhancement in turn produces that the function $T_{AA}^{e}$ is below 1 and
thus contributes to the suppression factor introduced by considering energy
losses due to the propagation of heavy flavors in the plasma. 

It is worth to keep in mind some important features concerning the results of
the present calculation: First, notice that we have not included the momentum
shift introduced by energy losses when computing the transverse distributions
of charmed mesons and 
baryons. This is so because for $R_{AA}^{e}$, energy losses should be included
in the prefactor of the function $T_{AA}^{e}$. In this sense, in order to
avoid a double counting of the effect, the ratio that goes into the
calculation of this last function is the {\it raw} ratio. Second, it is
expected that at some value of $p_T$, fragmentation becomes the dominant
mechanism for hadron production and therefore that the charmed baryon to meson
ratio decreases above that $p_T$ value, given that fragmentation produces
more mesons than baryons. Third, we have considered finite
values of transverse flow for charmed mesons and baryons even thought it might
be questionable that heavy flavors also flow as light flavors
do. Nevertheless, there seems to be some experimental support for heavy quark
flow~\cite{Butsyk}. In this sense, the flow strength range we have considered
is only for moderate values. Notice however that even in the absence of flow
the suppression factor keeps being less than 1. Some of these issues will be
the subject of a future work to appear elsewhere.

\section*{Acknowledgments}

A.A. wishes to thank the kind hospitality of both faculty and staff in
CBPF during a sabbatical visit. Support for this work has been
received in part by FAPERJ under Proj. No. E-26/110.166/2009, CNPq, the
Brazilian Council for Science and Technology, DGAPA-UNAM under PAPIIT grant
No. IN116008 and by CONACyT-M\'exico.

\end{document}